\documentclass[runningheads]{llncs}
\usepackage{graphicx}
\usepackage{adjustbox}
\usepackage{amsmath}
\usepackage{amssymb}
\usepackage{bbm}
\begin{document}
\title{Enhancing Single-Slice Segmentation with 3D-to-2D Unpaired Scan Distillation}
%
\titlerunning{3D-to-2D Unpaired Scan Distillation}
%
\author{Xin Yu\inst{1} \and
Qi Yang\inst{1} \and
Han Liu\inst{1} \and
Ho Hin Lee\inst{1} \and
Yucheng Tang\inst{2,3} \and
Lucas W. Remedios\inst{1} \and
Michael E. Kim \inst{1} \and
Rendong Zhang \inst{1} \and
Shunxing Bao \inst{2} \and
Yuankai Huo\inst{1} \and
Ann Zenobia Moore \inst{4}\and
Luigi Ferrucci \inst{4} \and
Bennett A. Landman\inst{1,2}}
\authorrunning{X. Yu et al.}
%
\institute{Computer Science, Vanderbilt University, Nashville, TN, USA \\
\and
Electrical and Computer Engineering,  Vanderbilt University, Nashville, TN, USA \and
Nivida Corporation \and
National Institute on Aging, Baltimore, MD, USA}
\maketitle              
\begin{abstract}

2D single-slice abdominal computed tomography (CT) enables the assessment of body habitus and organ health with low radiation exposure. However, single-slice data necessitates the use of 2D networks for segmentation, but these networks often struggle to capture contextual information effectively. Consequently, even when trained on identical datasets, 3D networks typically achieve superior segmentation results. In this work, we propose a novel 3D-to-2D distillation framework, leveraging pre-trained 3D models to enhance 2D single-slice segmentation. Specifically, we extract the prediction distribution centroid from the 3D representations, to guide the 2D student by learning intra- and inter-class correlation. Unlike traditional knowledge distillation methods that require the same data input, our approach employs unpaired 3D CT scans with any contrast to guide the 2D student model. Experiments conducted on 707 participants from the single-slice Baltimore Longitudinal Study of Aging (BLSA) dataset demonstrate that state-of-the-art 2D multi-organ segmentation methods can benefit from the 3D teacher model, achieving enhanced performance in single-slice multi-organ segmentation. Notably, our approach demonstrates considerable efficacy in low-data regimes, outperforming the model trained with all available training subjects even when utilizing only 200 training subjects. Thus, this work underscores the potential to alleviate manual annotation burdens.
\end{abstract}
%
%
\section{Introduction}

\begin{figure}[h]
 \includegraphics[width=\textwidth]{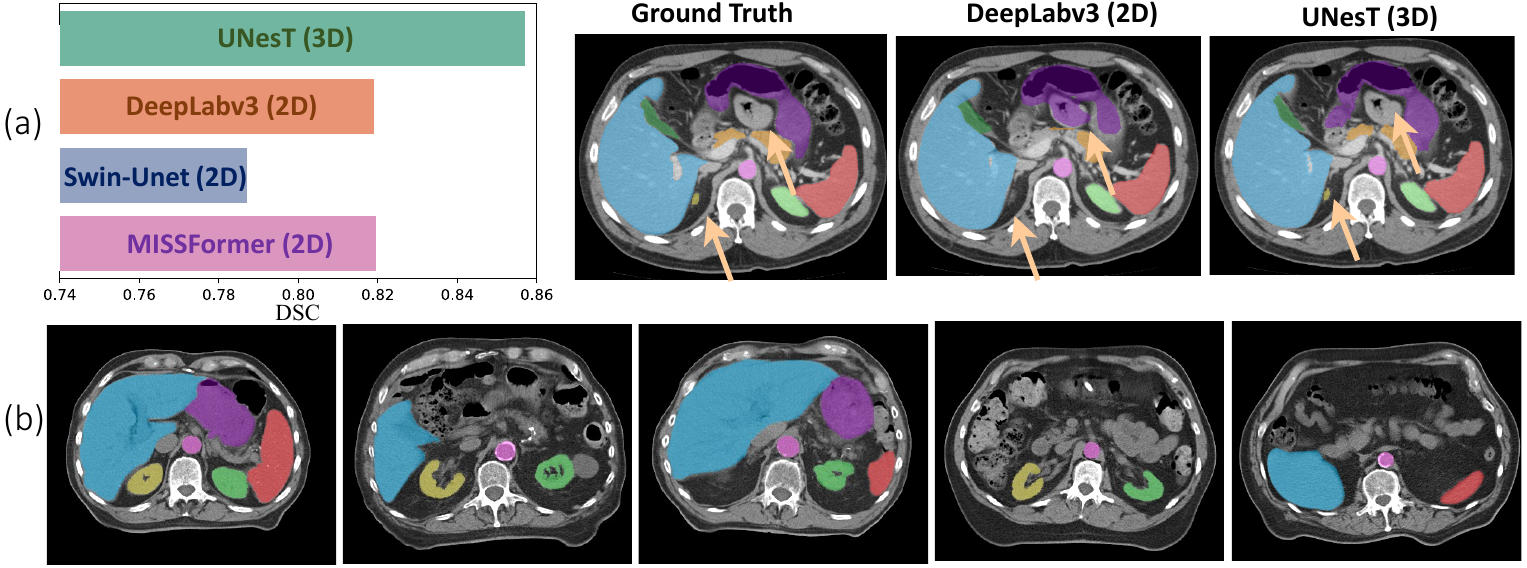}
\caption{(a) 3D and 2D model performance comparison. All the models are trained and tested on the MICCAI 2015 Multi-Atlas Abdomen Labeling Challenge (BTCV) dataset. UNesT~\cite{yu2023unest}, a 3D model, has superior performance (DSC) compared to other 2D models. (b) Example abdominal slices from the single-slice CT dataset with manual annotation on right/left kidneys, liver, spleen, stomach and aorta. The scans are in a wide range of vertebral levels. } 
\label{fig1}
\end{figure}

Abdominal computed tomography (CT) provides detailed tissue maps with high resolution that enables quantification of body composition. To better assess the internal structures and organ health, many deep learning-based multi-organ segmentation methods have been proposed and demonstrated their effectiveness~\cite{lee2023scaling,tang2022self,tang2021high,huang2022missformer,cao2022swin}. These methods can typically be categorized into two types: those trained with 2D slices (2D models) and those trained with 3D volumes (3D models). 2D models require lower machine memory costs while 3D methods benefit from more contextual information and usually lead to better segmentation results~\cite{liao2022deep,li2023transforming}, even when trained on identical data, as depicted in Fig.~\ref{fig1}(a). However, in routine monitoring and screening, 2D Non-contrast axial abdominal single-slice CT is preferred over 3D CT to minimize unnecessary radiation exposure~\cite{yu2022reducing}, necessitating the use of 2D networks. As the example single-slice shown in Fig.~\ref{fig1}(b), the heterogeneity in levels of image acquisition across subjects, along with the absence of three-dimensional context and contrast, makes the segmentation of single-slice with 2D models a challenge.

We aim to transfer the learned contextual knowledge from a pretrained 3D model to a 2D model to improve the performance on multi-organ segmentation. Knowledge distillation has been widely explored in the medical imaging domain. Most of the work has been focused on model compression by transferring knowledge from complex models to light-weight ones~\cite{Zhao2023EfficientMS,Elbatel2023FoProKDFP,qin2021efficient}. Some other works focus on using knowledge distillation to solve missing modality problem by distilling multi-modality knowledge to a single domain~\cite{wang2023prototype,wang2023learnable}. However, these works typically require the same or paired inputs for the teacher and student models since they require logit match of student and teacher output, which is not applicable in the single-slice dataset where paired 3D volumetric data is not available. Qou \textit{et al.}~\cite{Dou2020UnpairedMS} proposed an unpaired knowledge distillation framework for multi-modal segmentation tasks. However, in this work, the modality-specific knowledge is learned on a batch basis rather than a dataset basis, and the method is designed to handle different inputs of the same dimension. In the domain of natural image processing, many works have been proposed for 3D-to-2D knowledge transfer for indoor scene segmentation \cite{liu20213d,wu20223d,hou2021pri3d}. However, it is non-trivial to adapt these methods to our case due to the differences in image format as well as the requirement of paired 3D2D input, which is not applicable in our scenario. 


In this work, we aim to address two key challenges. Firstly, we aim to enable the 3D model to learn an effective embedding without requiring point-to-point matching when passing it to the 2D model. Secondly, we aim to facilitate the 2D model in learning from the embedding extracted from unpaired 3D data. To address the first challenge, we posit to extract the distribution centroid of each class using the 3D model and encourage the 2D centroid to be close to the 3D centroid during distillation. Inspired by \cite{liu2024cosst,zhang2021prototypical}, we compute the class-wise feature centroids (prototype) over the whole 3D dataset as the class centroid for the 3D model output and compute the sample-wise prototype on-the-fly for 2D samples. To address the second challenge, we extend the DIST loss introduced in \cite{huang2022knowledge} to maximize the linear inter- and intra-class correlation between 3D and 2D models instead of point-wise exact match used in conventional knowledge distillation. To this end, we design a novel unpaired 3D-to-2D knowledge distillation framework to improve 2D single-slice segmentation performance. We conduct extensive experiments on three representative 2D segmentation methods on 707 subjects from the single-slice Baltimore Longitudinal Study of Aging (BLSA) dataset with 3D model trained with CT data of different contrasts. The results demonstrate that our 3D-to-2D distillation framework can help improve 2D model segmentation.

Our contributions are three-fold: (1) to the best of our knowledge, we propose the first 3D-to-2D distillation framework in the medical imaging domain that leverages pretrained 3D models to improve 2D model segmentation; (2) our designed approach learns the 3D prototype during 3D pretraining, eliminating the need for 3D input during the inference phase; (3) we demonstrate that our proposed method consistently learns from the unpaired 3D prototype and enhances single-slice segmentation.

\section{Method}
\begin{figure}
\includegraphics[width=\textwidth]{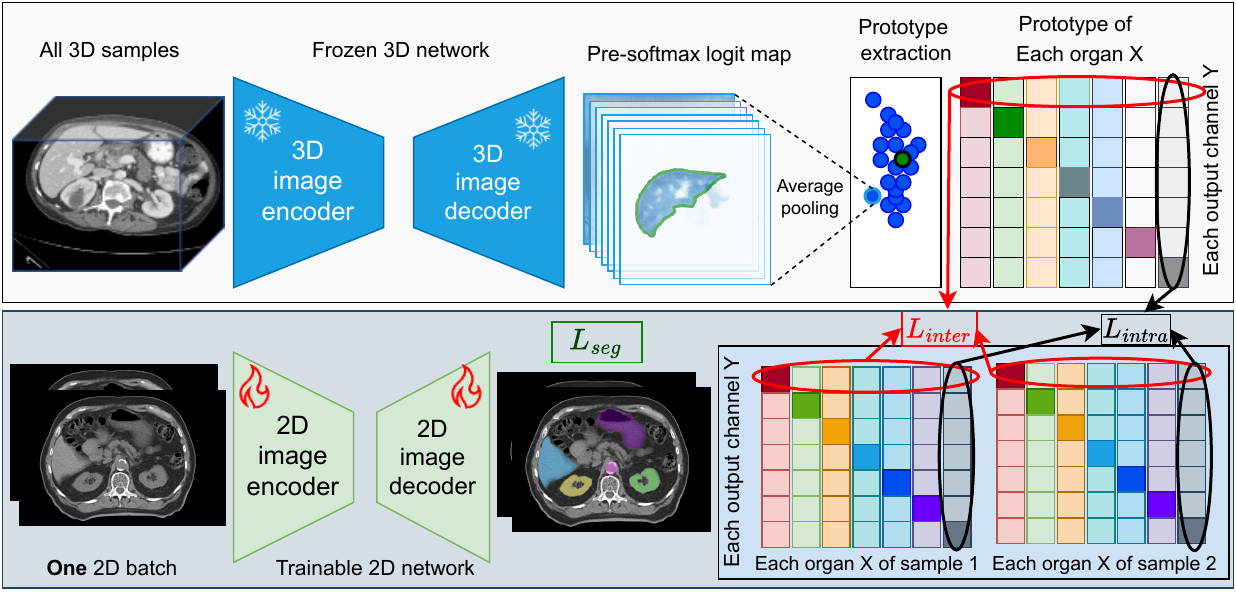}
\caption{The overview of the proposed method. The whole pipeline can be divided into a frozen 3D model (teacher) and a trainable 2D model (student). 3D model is pretrained to compute dataset class-wise feature centroid (prototype). 2D prototype is computed on-the-fly during distillation, and encouraged to be close to the 3D prototype via optimizing inter- and intra-class correlation. Each class prototype has the maximum response in the channel corresponding to its class index.} 
\label{fig2}
\end{figure}

The proposed 3D-to-2D distillation framework consists of a pretrained 3D segmentation network (teacher) and an active 2D segmentation network (student). The overall pipeline of the framework is shown in Fig.~\ref{fig2}. Before the distillation, both the 3D and 2D models are trained using training data before distillation. The 2D model is pretrained to obtain an initialization, whereas the 3D model is trained to acquire 3D features centroids (prototype).

\subsection{Prototype Extraction from 3D Model}
In our scenario, the aim is to enhance the segmentation performance of single-slice data using existing 3D datasets. Since the input 3D data and 2D data are unpaired, common pixel-wise distillation methods are inapplicable. We propose to align the 2D student model output to the distribution of the 3D teacher model output to preserve rich contextual information learned by 3D networks.

Inspired by \cite{liu2024cosst,zhang2021prototypical}, we propose computing the class-wise feature centroids vector (prototype) as the class distribution of the 3D model output. The underlying assumption for this approach is that the computed centroid is positioned closer to the true centroid of the underlying cluster \cite{zhang2021prototypical}.
For each slice $s$ in the 3D volume, we take the final feature map of the 3D network $\mathcal{X} \in \mathbb{R}^{H\times{W}\times{CH}}$ to calculate the prototype of each class, where $H$, $W$ and $CH$ represents height, width, and channel, respectively. The centroid of each organ $k$ in each slice $z_k$ is obtained by computing the average feature representation within the corresponding organ mask: 
\begin{equation}
z_k=\frac{\sum_i \mathcal{X}^i * \mathbbm{1}\left(y(i, k)==1\right)}{\sum_i \mathbbm{1}\left(y(i, k)==1\right)},
\end{equation}
where $\mathbbm{1}$ is the indicator function, and $i$, $y$ represent the $ith$ pixel in the given slice and ground truth, respectively. 

As the single-slice data primarily covers limited lumbar regions, we apply the body part regression method proposed in \cite{tang2021body} to crop the 3D volume and compute the prototype based on slices within this region. The final prototype of all classes has the size of $C \times CH \times 1 \times 1$, where $C$ and $CH$ represent class and channel, respectively.


\subsection{3D-to-2D distillation}
Given the prototype of 3D output, we calculate the 2D prototype of 2D slices input on-the-fly during distillation. To distill the learned prototype from the 3D model to the 2D student, we extend the DIST loss introduced in \cite{huang2022knowledge}. This loss function aims to preserve both inter-class and intra-class correlation between the student and teacher models. This is in contrast to conventional knowledge distillation methods that aim to align the student's outputs with the teacher's output on a point-wise basis, which is not applicable in our unpaired cases. 
In our case, each class is represented by a vector as opposed to a value in the original form. We encourage the class-wise prototype from the student $X^{\mathrm{s}}$ to be close to the teachers' $X^{\mathrm{t}}$, as shown in Fig.~\ref{fig2} by optimizing the inter-class loss, written as:

\begin{equation}
\mathcal{L}_{\text {inter }}=\frac{1}{B} \sum_{i=1}^B d\left(\boldsymbol{X}_{i}^{\mathrm{s}}, \boldsymbol{X}^{\mathrm{t}}\right),
\end{equation}

where $B$ represents batch size,  $d(x,y) = 1 - \rho(x,y)$, and $\rho(x,y)$ represents the output of Pearson’s correlation coefficient.

Since we only have one teacher prototype per dataset rather than a batch, instead of calculating the intra-class loss based on the batch as done in the original form, we compute the intra-class loss based on the channel vector of each class, as shown in Fig.~\ref{fig2}. This is because we use the final feature map to calculate the class prototype; we believe that the channel vector contains responses from other classes regarding this class. We extend the loss to encourage the student channel $Y^s$ to be close to the teacher channel distribution $Y^t$ by optimizing the intra-class loss written as:

\begin{equation}
\mathcal{L}_{\text {intra }}=\frac{1}{B} \sum_{i=1}^B d\left(\boldsymbol{Y}_{i}^{\mathrm{s}}, \boldsymbol{Y}^{\mathrm{t}}\right)
\end{equation}
The total extended DIST loss with inter- and intra-loss can be written as: 
\begin{equation}
\mathcal{L}_{\text {DIST}} = \mathcal{L}_{\text {inter}} +  \mathcal{L}_{\text {intra}}
\end{equation}
Together with the original segmentation loss, the final loss function for the 3D-to-2D distillation is written as: 
\begin{equation}
    \mathcal{L} = \mathcal{L}_{\text {seg }} + \beta\mathcal{L}_{\text {DIST}}
\label{eq4}
\end{equation}
where $\beta$ is a weighting factor to balance different losses. 

The 3D contextual information is implicitly learned during training and no 3D prototype is required during inference.



\section{Experiments and Results}
\subsection{Dataset}
The 3D teacher models are trained with 3D volumetric CT datasets in both the Non-contrast phase and the Portal Venous phase, and the 2D student models are trained with BLSA Non-contrast single-slice dataset.

\textbf{3D dataset} The 3D Non-contrast dataset is an in-house 3D dataset collected on 49 splenomegaly subjects with expert-refined annotations for the major organs. We use 30 Portal Venous CT volumes from the MICCAI 2015 Multi-Atlas Abdomen Labeling Challenge dataset (BTCV) to evaluate the case when 3D training data is of different contrast than the single-slice dataset. 

\textbf{Single-slice dataset} We use 796 single slices from 707 subjects to train and evaluate the 2D models. The dataset is split into training, validation, and test sets with 295, 117, and 295 different subjects, respectively. All slices are Non-contrast axial slices in the abdominal region containing at least one organ. Due to the lack of context and contrast in the non-contrast single-slice dataset, only major organs including the spleen, right/left kidneys, liver, stomach, and aorta are annotated.

\subsection{Implementation Details}
We use UNesT \cite{yu2023unest} as the 3D teacher model, as it proved effective on learning 3D medical representation. We use 2D representative models for both natural image processing and multi-organ segmentation DeepLabv3 \cite{chen2017rethinking}, Swin-Unet \cite{cao2022swin}, and MISSFormer \cite{huang2022missformer} as our student models. The teacher model is pretrained with the 3D dataset, and the student models are pretrained with the single-slice dataset following the multi-organ segmentation training strategy mentioned in the original paper. For DeepLabv3 however, we adopt the settings outlined in \cite{yu2023longitudinal}, which are more suitable for multi-organ segmentation. Once the 3D model is trained, we store the feature maps from the final layer of the model to compute the prototype offline.  After student model has converged, we select the best model on the validation set as the initialization model for distillation, ensuring that performance gains are solely attributed to distillation rather than continued training. During the 3D-2D distillation process, we set the value of $\beta$ in Equation~\ref{eq4} to 0.5 to balance its contribution with the segmentation loss. The distillation process is trained for 100 epochs.

\begin{table}[]
\caption{The performance comparison between different methods before and after distillation with the 3D model trained on different contrast CT data. Note: RKidney: right kidney, LKidney: left kidney. }
\label{table1}
\begin{adjustbox}{width=\textwidth}
\begin{tabular}{lccccccc}
\hline
\multicolumn{1}{l|}{Methods}            & Spleen         & RKidney        & LKidney        & Liver          & Stomach        & \multicolumn{1}{c|}{Aorta}          & Avg            \\ \hline\hline
\multicolumn{1}{l|}{DeepLabv3}         & 0.769          & 0.826          & 0.775          & 0.878          & 0.768          & \multicolumn{1}{c|}{0.802}          & 0.806          \\
\multicolumn{1}{l|}{Swin-Unet}          & 0.779          & 0.846          & 0.850          & 0.877          & 0.744          & \multicolumn{1}{c|}{0.761}          & 0.811          \\
\multicolumn{1}{l|}{MISSFormer}         & 0.782          & 0.865          & \textbf{0.870} & 0.884          & 0.773          & \multicolumn{1}{c|}{0.832}          & 0.841          \\ \hline
\multicolumn{8}{c}{distilled with UNesT trained with the BTCV dataset}                                                                                                            \\ \hline
\multicolumn{1}{l|}{DeepLabv3 distill} & 0.753          & 0.828          & 0.828          & 0.888          & 0.813          & \multicolumn{1}{c|}{0.799}          & 0.825          \\
\multicolumn{1}{l|}{Swin-Unet distill}  & 0.755          & 0.850          & 0.856          & 0.875          & 0.769          & \multicolumn{1}{c|}{0.799}          & 0.824          \\
\multicolumn{1}{l|}{MISSFormer distill} & \textbf{0.816} & 0.865          & 0.860          & 0.886          & 0.783          & \multicolumn{1}{c|}{0.829}          & 0.842          \\ \hline
\multicolumn{8}{c}{distilled with UNesT trained with in-house non-contrast dataset}                                                                                               \\ \hline
\multicolumn{1}{l|}{DeepLabv3 distill} & 0.752          & 0.843          & 0.817          & \textbf{0.897} & \textbf{0.825} & \multicolumn{1}{c|}{0.778}          & 0.824          \\
\multicolumn{1}{l|}{Swin-Unet distill}  & 0.779          & 0.845          & 0.857          & 0.879          & 0.754          & \multicolumn{1}{c|}{0.784}          & 0.821          \\
\multicolumn{1}{l|}{MISSFormer distill} & 0.805          & \textbf{0.870} & \textbf{0.870}          & 0.886          & 0.769          & \multicolumn{1}{c|}{\textbf{0.838}} & \textbf{0.845} \\ \hline
\end{tabular}
\end{adjustbox}
\end{table}

\subsection{Results}

\begin{figure}
\includegraphics[width=\textwidth]{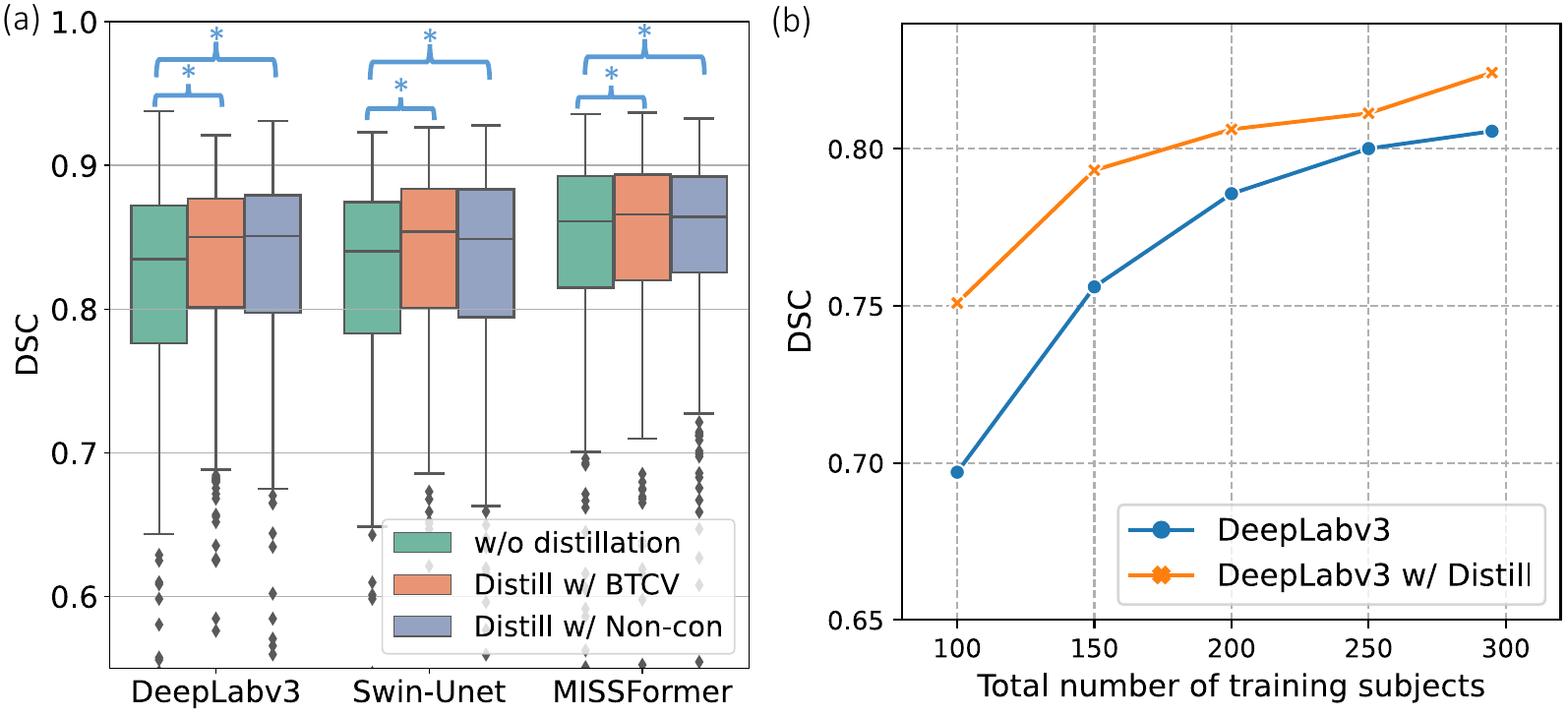}
\caption{Comparison of segmentation of models trained with and without distillation. In (a), models with distillation consistently reduce the variation with improved median and quartiles. * indicates statistically significant ($p < 0.05$) by Wilcoxon signed-rank test. (b) shows the results of DeepLabv3 w/ and w/o distillation using different number of training subjects.  } 
\label{fig3}
\end{figure}

\begin{figure}
\includegraphics[width=\textwidth]{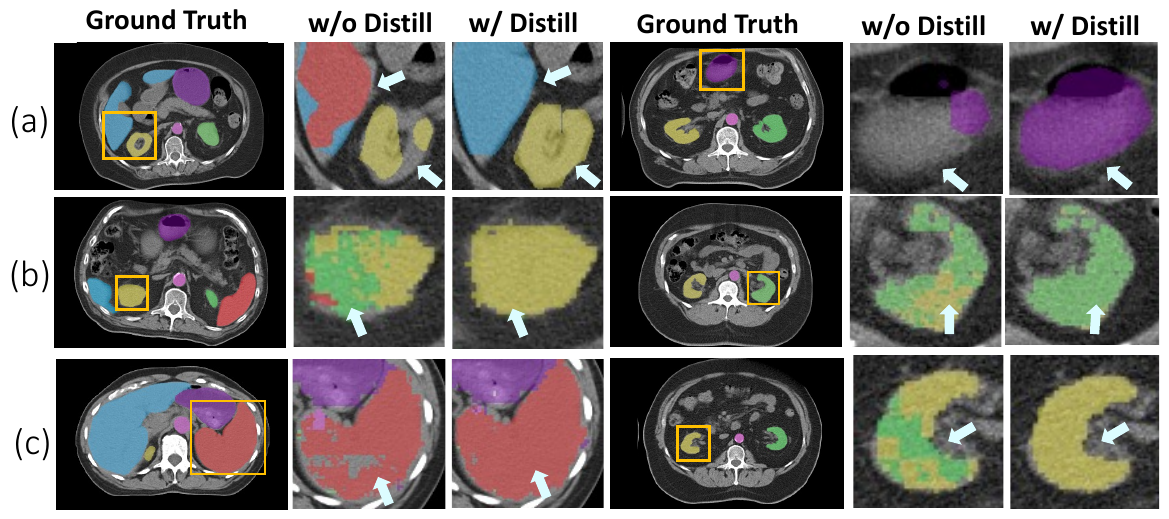}
\caption{Qualitative results of models with and without distillation. (a), (b), and (c) represent the results of DeepLabv3, Swin-Unet and MISSFormer, respectively. White arrows emphasized the segmentation improvement on right kidney (yellow), left kidney (green), spleen (red), liver (blue), and stomach (purple).} 
\label{fig4}
\end{figure}

We present the quantitative results in Table~\ref{table1} and Fig.~\ref{fig3}, and qualitative segmentation in Fig.~\ref{fig4}. We use Dice Similarity Coefficient (DSC) as the metric to evaluate model performance. Comparing results before and after 3D-to-2D distillation, the results after distillation consistently outperform the results without distillation significantly with $p < 0.05$ under the Wilcoxon signed-rank test, especially on DeepLabv3 where distillation improves the results by 2.4\%. 

As shown in Fig.~\ref{fig3}(a), distillation further reduces the performance variation with improved median and tighter quartiles. As shown in Fig~\ref{fig4} and Table~\ref{table1}, most of the models achieve large improvement on right/left kidneys where distillation helps models to distinguish right and left kidneys and improve the under-segmentation problem. The DSC of the spleen shows the largest fluctuation among different models, primarily because only 111 out of 348 slices in the test set contain the spleen. This may explain why MISSFormer exhibits improvement in most organs, especially in the spleen, but less improvement on average. Comparing the distillation results with the 3D teacher model trained on both the Portal Venous and Non-contrast CT datasets, the models achieve comparable performance across different student models, indicating that the method does not require the same input contrast type.

\noindent {\bf Low-data regime analysis}
 We investigate the 3D-to-2D distillation's effectiveness in low-data regime using DeepLabv3 model with 3D teacher pretrained with Non-contrast dataset. Fig.~\ref{fig3}(b) shows the performance comparison between models trained on 100, 150, 200, 250 and all subjects in the training set. 
The model consistently outperforms the original models after distillation. We observe that the model trained with 200 subjects after distillation achieves better performance than the model trained without distillation, using all available training samples.
\section{Discussions and Conclusions}
As the first work of leveraging 3D embedding to assist 2D segmentation in the medical imaging domain, our work has many limitations. We only explore our method on the major/large organs where the performance gap between 2D and 3D models is not particularly large, as the single-slice dataset currently only has annotations for major organs. Based on the experimental observation, the method can only enhance the original 2D network performance to a certain extent and cannot bring it close to the performance of 3D models. We exclusively leverage 3D features learned by UNesT in our experiments, lacking additional 3D models for a comprehensive evaluation. Moreover, the generalizability of our method can be further explored by pretraining it with diverse 3D datasets, such as MRI, to assess its performance across different training data modalities.

In this paper, we introduce a novel 3D-to-2D distillation framework designed to enhance 2D single-slice segmentation by integrating embeddings derived from 3D models trained on 3D datasets. The model specifically learns to align the inter-class and intra-class correlation during training and requires no 3D input during inference. Experiments on single-slice dataset with three representative 2D segmentation models demonstrate the effectiveness of our framework. Moving forward, we hope our work could inspire future endeavors aimed at integrating 3D information to enhance 2D segmentation techniques.

\noindent {\bf Acknowledgements}
This research is supported by NSF CAREER 1452485, 2040462 and the National Institutes of Health (NIH) under award numbers R01EB017230, R01EB006136, R01NS09529, T32EB001628, 5UL1TR002243-04, 1R01MH121620-01, and T32GM007347; by ViSE/VICTR VR3029; and by the National Center for Research Resources, Grant UL1RR024975-01, and is now at the National Center for Advancing Translational Sciences, Grant 2UL1TR000445-06. This research was conducted with the support from the Intramural Research Program of the National Institute on Aging of the NIH. The content is solely the responsibility of the authors and does not necessarily represent the official views of the NIH. The identified datasets used for the analysis described were obtained from the Research Derivative (RD), database of clinical and related data. The inhouse imaging dataset(s) used for the analysis described were obtained from ImageVU, a research repository of medical imaging data and image-related metadata. ImageVU and RD are supported by the VICTR CTSA award (ULTR000445 from NCATS/NIH) and Vanderbilt University Medical Center institutional funding. ImageVU pilot work was also funded by PCORI (contract CDRN-1306-04869).

%
%
%
%
\bibliographystyle{splncs04}
\bibliography{reference}




\end{document}